\documentclass[%
superscriptaddress,
preprint,amsmath,amssymb, aps,pra,
]{revtex4-2}

\usepackage[english]{babel}
\usepackage{amsmath}
\usepackage{graphicx}
\usepackage{float}
\usepackage{subcaption}
\usepackage{dcolumn}% Align table columns on decimal point
\usepackage{bm}% bold math
\usepackage[colorlinks=true, allcolors=blue]{hyperref}

\usepackage[version=4]{mhchem}
\usepackage[table,xcdraw]{xcolor}  % in your preamble

\newcommand{\doHMN}[2]{%
  \begingroup\lccode`~=`#1
  \lowercase{\endgroup\let~}#2%
  \mathcode`#1="8000}

\definecolor{myblue}{RGB}{235,245,255} 
\begin{document}

\title{Atomistic modeling of uranium monocarbide with a machine learning interatomic potential}

\author{Lorena Alzate-Vargas}
\affiliation{Theoretical Division, Los Alamos National Laboratory, Los Alamos, New Mexico 87545, USA}%
\author{Kashi N. Subedi}
\affiliation{Theoretical Division, Los Alamos National Laboratory, Los Alamos, New Mexico 87545, USA}%
\author{Roxanne M. Tutchton}
\affiliation{Theoretical Division, Los Alamos National Laboratory, Los Alamos, New Mexico 87545, USA}%
\author{Michael W.D. Cooper}
\affiliation{Materials Science and Technology Division, Los Alamos National Laboratory, Los Alamos, New Mexico 87545, USA}%
\author{Tammie Gibson}
\affiliation{Theoretical Division, Los Alamos National Laboratory, Los Alamos, New Mexico 87545, USA}%
\author{Richard A. Messerly}
\affiliation{Theoretical Division, Los Alamos National Laboratory, Los Alamos, New Mexico 87545, USA}%
\affiliation{National Center for Computational Sciences Division, Oak Ridge National Laboratory, Oak Ridge, TN 37830, USA}%

\begin{abstract}
Uranium monocarbide (UC) is an advanced ceramic fuel candidate due to its superior uranium density and thermal conductivity compared to traditional fuels. To accurately model UC at reactor operating conditions, we developed a machine learning interatomic potential (MLIP) using an active learning procedure to generate a comprehensive training dataset capturing diverse atomic configurations. The resulting MLIP predicts structural, elastic, thermophysical properties, defect formation energies, and diffusion behaviors, aligning well with experimental and theoretical benchmarks. This work significantly advances computational methods to explore UC, enabling efficient large-scale and long-time molecular dynamics simulations essential for reactor fuel qualification.
\end{abstract}

\maketitle

\vfill
\begin{center}
\footnotesize{
This manuscript has been authored in part by UT-Battelle, LLC, under contract DE-AC05-00OR22725 with the U.S. Department of Energy (DOE). The U.S. government retains and the publisher, by accepting the article for publication, acknowledges that the U.S. government retains a nonexclusive, paid-up, irrevocable, worldwide license to publish or reproduce the published form of this manuscript, or allow others to do so, for U.S. government purposes.  DOE will provide public access to these results of federally sponsored research in accordance with the DOE Public Access Plan (http://energy.gov/downloads/doe-public-access-plan).}
\end{center}

\section{Introduction}

Uranium monocarbide (UC) is an advanced ceramic fuel considered promising for use in Generation-IV nuclear reactor systems, such as gas-cooled and sodium-cooled fast reactors, due to its favorable nuclear and thermal properties~\cite{konovalov2003, Petti2009}. Specifically, UC possesses a significantly higher uranium density and superior thermal conductivity compared to conventional oxide-based fuels like uranium dioxide (UO$_2$), making it attractive from both an economic and a safety perspective~\cite{matzke1986, Frost1963, Vasudevamurthy2022}. However, these advantageous properties are accompanied by critical challenges, including pronounced fuel swelling and substantial fission gas release under operational conditions, complicating its practical deployment.

To overcome these challenges and evaluate the viability of UC as a nuclear fuel, a robust understanding of the underlying structure-property relationship is essential. However, experimental research on nuclear fuel materials presents many difficulties as it requires processing and measuring radioactive samples. Atomistic modeling methods, particularly density functional theory (DFT), have proven to be effective in studying fundamental properties of UC such as point defect formation, diffusion mechanisms, and incorporation of fission products within nuclear fuels~\cite{Freyss2010,Ducher2011,Bevillon2012,Bevillon2013}.

Previous work~\cite{Freyss2010} reported point defect formation energies and provided valuable insights into the accommodation of helium, xenon, and oxygen impurities. Later work~\cite{Ducher2011} further investigated defect migration mechanisms, revealing that uranium and carbon vacancies play a dominant role in atomic diffusion processes. Additional studies~\cite{Bevillon2012, Bevillon2013} expanded on these results by analyzing the incorporation energies and diffusion behaviors of fission products, highlighting the efficacy of DFT in capturing fundamental defect phenomena.

However, DFT has notable limitations when applied to strongly correlated materials such as UC, particularly due to difficulties in accurately representing the localized nature of the uranium 5\textit{f} electrons. To better account for electron-electron interactions in strongly correlated electron systems, one can improve the Coulomb interaction term in DFT calculations by adding the the Hubbard (\textit{U}) parameter. DFT+\textit{U} markedly improves the prediction of elastic constants, and electronic structure of UC~\cite{Shi2009}. Similarly, it gives accurate lattice dynamics, phonon dispersions, and overall thermodynamic stability~\cite{Wdowik2016} compared to DFT, bringing the computational results closer in alignment with the experimental observations.

More recent work~\cite{Huang2020} specifically utilized DFT+\textit{U} calculations to study intrinsic defects and xenon impurities, providing detailed energetic insights essential for understanding irradiation behavior. These findings underscore that the Hubbard term provides more reliable data on defect energetics and diffusion barriers, crucial parameters for accurately modeling UC under reactor operating conditions.

Although computing certain single-point properties with DFT+\textit{U} is relatively straightforward, predicting structural and thermophysical properties requires long-time averaging of molecular dynamics (MD) simulations. Despite advancements in electronic structure methods and computing hardware, performing long MD simulations with DFT+\textit{U} remains extremely computationally expensive. Classical potentials provide a more practical route from a computational standpoint, however existing classical potentials exhibit only moderate accuracy in reproducing experimental and theoretical results for structural and thermophysical properties~\cite{Basak2007, Chartier2007}. Therefore, there is a clear need for computationally efficient yet highly accurate approaches to effectively model UC at large scales and extended timescales relevant to reactor conditions.

Machine learning interatomic potentials (MLIPs) offer a promising solution to bridge this gap, combining near-DFT+\textit{U} accuracy with computational efficiencies that remain comparable to classical potentials. Recent implementations of MLIPs have shown remarkable success in accurately modeling uranium-based materials, including elemental uranium~\cite{Chen2023}, uranium dioxide (UO$_2$)~\cite{Stippell2024,Dubois2024} and uranium mononitride (UN)~\cite{alzatevargas2024}, across a wide range of thermophysical conditions and defect configurations. These studies have demonstrated the importance of active learning for generating a diverse and representative training dataset. A properly trained MLIP enables detailed exploration of nuclear fuels under realistic operating conditions.

In this work, we present the first MLIP specifically developed for uranium monocarbide, based on the Hierarchically Interacting Particle Neural Network (HIP-NN) architecture~\cite{Lubbers2018,Chigaev2023}. We utilize an active learning framework guided by uncertainty quantification and MD sampling to generate a DFT+\textit{U} quality training dataset for UC, which includes various defect structures relevant to nuclear applications of UC. We subsequently validate our MLIP by predicting fundamental thermophysical properties, elastic constants, defect energies, migration and self-diffusion activation energies, comparing these predictions against experimental data and DFT/DFT+\textit{U} calculations. In the absence of reliable classical potentials or experimental data, our MLIP provides predictions for properties that were previously unattainable with DFT+\textit{U}. Our results demonstrate that the developed MLIP reliably captures critical features of UC, significantly advancing our ability to perform comprehensive and accurate atomistic simulations of this advanced nuclear fuel.

\section{Methods}

\subsection{Density functional theory calculations}

All density functional theory (DFT) calculations performed in this study used the frozen-core, all-electron projector-augmented-wave (PAW) method~\cite{Bloch1994}, implemented in the Vienna \textit{ab initio} simulation package (VASP)~\cite{Kresse1996,Kresse1999}. We used the generalized gradient approximation (GGA) as parametrized by Perdew, Burke, and Ernzerhof (PBE)~\cite{Perdew1996} as the exchange-correlation functional. A plane-wave basis set with a cutoff energy of 520 eV was used, along with the Methfessel-Paxton smearing scheme (smearing width of 0.1 eV) for electronic occupancies. The energy and force convergence criteria were set as $10^{-8}$ eV and $10^{-4}$ eV/\AA, respectively.

To accurately describe the strongly correlated nature of uranium 5$f$ electrons, we incorporated the DFT+\textit{U} approach developed by Dudarev et al.~\cite{Dudarev_1998}. Within this method, only the effective Hubbard parameter ($U_{\mathrm{eff}}=U-J$), representing the difference between the Hubbard parameter $U$ and the exchange parameter $J$, is explicitly defined.

\begin{figure}[ht!]
    \centering 
    \includegraphics[width=0.4\linewidth]{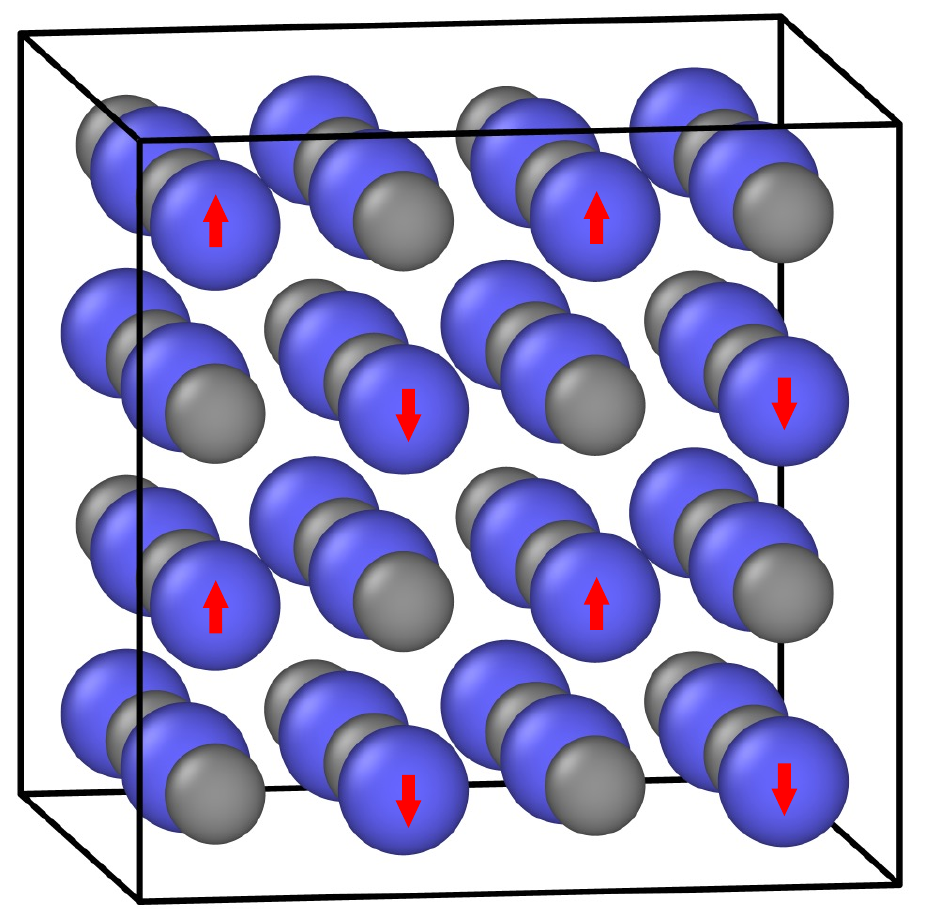} 
  \caption{\label{fig:cell} Perspective view of the UC crystal structure in the AFM-I phase and magnetic moments for uranium along the (001) direction (red arrows). The U atoms and C atoms are represented by blue and gray spheres, respectively.}
\end{figure}

We carried out a systematic investigation to determine the optimal value of $U_{\mathrm{eff}}$, exploring both non-magnetic (NM) and type-I antiferromagnetic (AFM-I) orderings in a NaCl-type structure. The AFM ordering was modeled by assigning alternating magnetic moments to uranium atoms in consecutive (001) planes, resulting in a net zero magnetic moment, as depicted in Fig.~\ref{fig:cell}. We performed calculations for various $U_{\mathrm{eff}}$ values, keeping the exchange parameter fixed at $J = 0.5$ eV. For each value of $U_{\mathrm{eff}}$, the lattice parameter, elastic constants, and phonon dispersions were computed and compared against experimental data.

Unit cell calculations of lattice parameter and elastic constants were performed using a dense $12 \times 12 \times 12$ Monkhorst-Pack $k$-point mesh with Brillouin zone integration. The input files required for elastic constant calculations were generated using VASPKIT~\cite{VASPKIT}. Phonon dispersion curves were calculated with the PHONOPY code~\cite{phonopy-phono3py-JPSJ}, utilizing the finite displacement method with atomic displacements set at 0.015 \AA. Phonon calculations used a $2 \times 2 \times 2$ supercell containing 64 atoms and a $4 \times 4 \times 4$ k-point mesh.

\begin{figure}[ht!]
  \centering
  \begin{subfigure}{0.48\textwidth}
    \includegraphics[width=\textwidth]{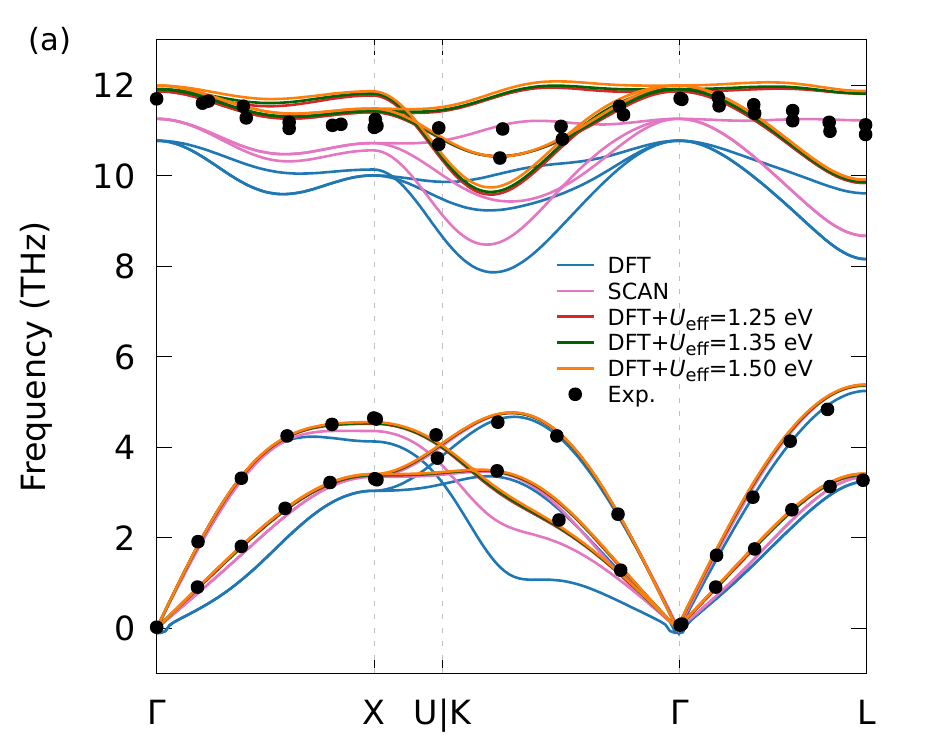}
    \label{fig:nm}
  \end{subfigure}
  \hspace{0.01\textwidth}
  \begin{subfigure}{0.48\textwidth}
    \includegraphics[width=\textwidth]{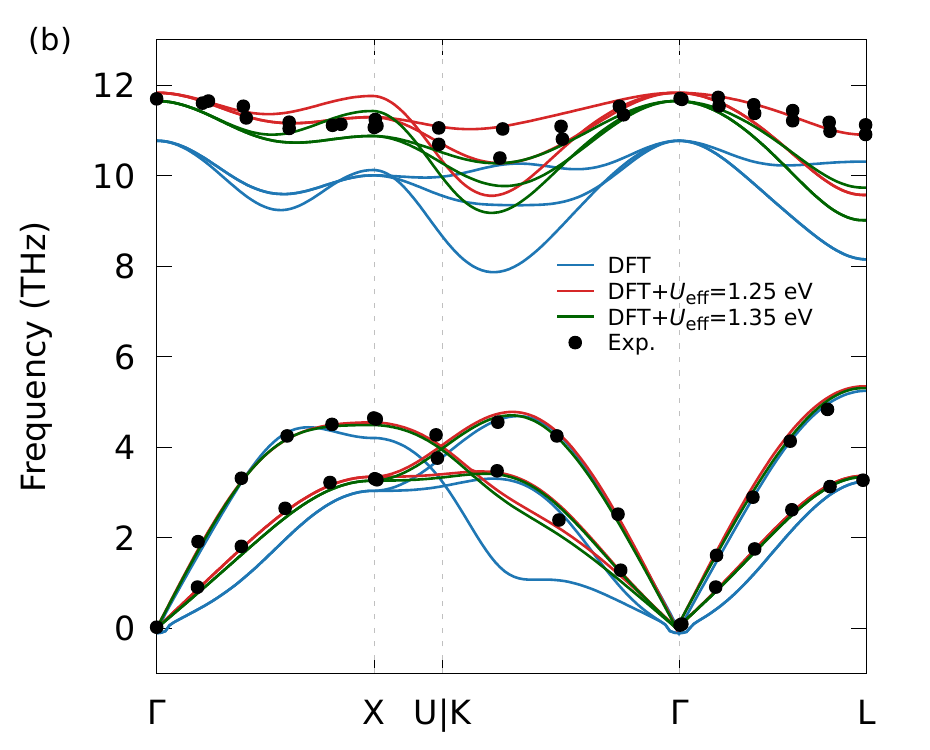}
    \label{fig:afm}
  \end{subfigure}
  \vspace{-20pt}
  \caption{Phonon band structures for (a) NM ordering and (b) AFM ordering using DFT, SCAN and DFT+\textit{U} with different $U_{\mathrm{eff}}$ values. Experimental data from Ref.~\cite{jackman1986} included as black points. AFM structure with $U_{\mathrm{eff}}$=1.25 eV calculations capture experimental optical modes better than the other values.}
  \label{fig:phonon}
\end{figure}

The computed lattice constants, bulk moduli, and elastic constants for different $U_{\mathrm{eff}}$ values and magnetic configurations are summarized in Table~\ref{tab:dft}. We also included the corresponding values when using DFT on both configurations and SCAN on the NM ordering, where we can see a large deviation in the prediction of $C_{11}$ and $C_{44}$.
However, when comparing the DFT+\textit{U} predictions, no single choice of $U_{\mathrm{eff}}$ simultaneously reproduces all experimental properties perfectly. Generally, NM ordering calculations tend to overestimate the elastic constant $C_{12}$ and bulk modulus despite reasonable predictions of the lattice parameter. Conversely, AFM ordering systems with $U_{\mathrm{eff}}$ values of 1.25 and 1.35 eV exhibit improved consistency with experimental data. Among these, $U_{\mathrm{eff}} = 1.25$ eV provides the closest agreement with experimental phonon dispersions (shown in Fig.~\ref{fig:phonon}) and bulk moduli. Consequently, we selected a $U_{\mathrm{eff}}$ value of 1.25 eV in the AFM-I configuration for performing all reference DFT+\textit{U} calculations in this study.

\begin{table*}[ht!] 
\caption{\label{tab:dft} Equilibrium properties of UC, including lattice
parameter, bulk modulus, and elastic constants calculated using different $U_{\mathrm{eff}}$ values for DFT+\textit{U} at 0~K and comparing NM and AFM magnetic orderings.}
\begin{ruledtabular}
\begin{tabular}{ccccccc}

Functional & Ordering  & Lattice parameter  & $B$  &$C_{11}$  & $C_{12}$ & $C_{44}$  \\
                & &  (\AA) & (GPa) & (GPa) & (GPa) &  (GPa) \\
\hline
DFT & NM &  4.933 & 187.1 & 290.5 & 135.3 & 24.7 \\
DFT & AFM & 4.933 & 186.7 & 291.5 & 134.4 & 24.7 \\
SCAN & NM & 4.938 & 181.7 & 290.7 & 132.0 & 33.3 \\
DFT+\textit{U} ($U_{\mathrm{eff}}$=1.25 eV)  & NM  & 4.950 & 191.5 & 304.8 & 134.8 & 43.2 \\
%\rowcolor{blue!6}[1.2ex]
\rowcolor{myblue}[1.2ex]
DFT+\textit{U} ($U_{\mathrm{eff}}$=1.25 eV)  & AFM & 4.953 & 159.4 & 332.7 & 72.8  & 47.9 \\
DFT+\textit{U} ($U_{\mathrm{eff}}$=1.35 eV)  & NM  & 4.951 & 192.1 & 306.3 & 135.0 & 44.7 \\
DFT+\textit{U} ($U_{\mathrm{eff}}$=1.35 eV)  & AFM & 4.957 & 151.4 & 318.5 & 67.9  & 50.4 \\
DFT+\textit{U} ($U_{\mathrm{eff}}$=1.50 eV)  & NM  & 4.953 & 192.8 & 308.4 & 135.0 & 47.0 \\
Exp. at 300~K (Refs.~\cite{graham1963,routbort1971,Butorin2023}) & -   & 4.956 & 156--158 & 314--320 & 77--79   & 64--66 \\
\end{tabular} 
\end{ruledtabular}
\end{table*}

\subsection{Active learning procedure for dataset generation}

An accurate MLIP depends on the diversity and representativeness of its training dataset. To systematically and efficiently construct such, we employed an active learning (AL) methodology proposed in Ref.~\cite{smith2018,Smith2021}.
The AL procedure aims to identify high-uncertainty atomic configurations by quantifying the variance in predicted energies and forces across an ensemble of eight semi-independent MLIPs. Each MLIP in the ensemble was trained on distinct random splits of training (80\%), validation (10\%), and test (10\%)) datasets and initialized with different model parameters. Configurations with high variance (high uncertainty) among the ensemble predictions were selected for inclusion in subsequent iterations of training, enhancing model accuracy and robustness. New candidate structures were sampled using MD simulations driven by an individual member of the ensemble of MLIPs. To facilitate rapid exploration of the potential energy surface, the temperature and density were dynamically modulated during each MD simulation through combined linear and sinusoidal adjustments, following work done in Ref.~\cite{Smith2021}. Temperatures were varied from an initial temperature between 1000--5000 K to a final temperature between 100--5000 K, while densities were selected between 10 and 17 g/cm$^3$. Density variations were implemented by scaling both the atomic positions and simulation cell dimensions accordingly, allowing the system to explore configurations representative of realistic operational scenarios and defect states.

The AL loop was initiated from a bootstrap dataset consisting of 200 configurations which were carefully selected from a subset of the training dataset previously developed using AL for uranium nitride MLIP~\cite{alzatevargas2024}. In this work, each N atom was substituted with a C atom.
The bootstrap dataset includes face-centered cubic NaCl-type structures, high-temperature systems, and various defect configurations (e.g., Frenkel pairs and C/U bivacancies). Given the small difference between the lattice parameters of UC (recommended value of 4.96~\AA~\cite{Butorin2023}, as value varies from 4.91--4.99~\AA~\cite{Vasudevamurthy2022} depending on carbon content,) and UN (4.88~\AA), we rescaled the atomic positions and simulation boxes for the initial configurations by approximately 1.4\% to better represent the equilibrium U-C bond lengths.

Single-point DFT+\textit{U} calculations were performed to compute energies and atomic forces for each configuration selected by AL. As a compromise between computational cost and efficiency, a $k$-point mesh of $2 \times 2 \times 2$ was adopted.

The final training dataset consisted of 6,742 configurations of $2 \times 2 \times 2$ supercells containing approximately 64 atoms, with slight variations due to defect formation, providing diverse structures and comprehensive coverage of relevant atomic environments required for a robust and reliable MLIP. 

\subsection{Machine learning potential}

HIP-NN is a message-passing neural network where the characterization of the environment takes place through learned parameters using sensitivity functions which characterize distances between atoms, and interaction functions which combine the distances and weights of neighboring atoms to form a message. HIP-NN was successfully implemented to model uranium nitride over a wide temperature range~\cite{alzatevargas2024}. Therefore, we used the same network parameters selected for UN, i.e, one interaction layer ($n_\textrm{interaction}=1$), $n_\textrm{feature}=60$ neurons per layer and $n_\nu=20$ sensitivity functions. 

To enhance the representation of angular correlations, we further employ the tensor-sensitive HIP-NN architecture~\cite{Chigaev2023} with order ${\ell_\text{max}}=1$ (HipnnVec), corresponding to vector-based message passing, consistent with previous work for UN. In this formulation, messages from neighboring atoms are accumulated not only as scalars but also as vectors, enabling the network to encode directional information and sensitivity to angles between neighbors. For completeness, supplementary information compares the HipnnVec model performance with that of a HIP-NN model trained with higher order ${\ell_\text{max}}=2$ tensor sensitivity to include quadrupole information (HipnnQuad).

For tensor-sensitive models, a cusp regularization is applied to stabilize the norm used to reduce vector features in highly symmetric local environments. In HIP-NN, vector features may vanish near equilibrium configurations, leading to poorly conditioned derivatives of the learned potential. Cusp regularization introduces a small positive constant inside the tensor norm, enforcing smooth first- and second-order derivatives with respect to atomic displacements. We have opted for a cusp value of $10^{-2}$ to balance harmonic stability against transferability to defected environments. The ensemble of eight semi-independent MLIPs was trained using \textit{hippynn}, the open-source implementation of HIP-NN~\cite{hippynn}.

\subsection{Molecular dynamics simulations}

MD simulations to validate and evaluate the performance of the developed MLIP were performed using the Large-scale Atomic/Molecular Massively Parallel Simulator (LAMMPS) software package~\cite{Thompson2022}, which directly interfaces with the HIP-NN architecture.

To characterize the mechanical response of uranium monocarbide at zero Kelvin, we computed the elastic constants using the deformation method. These calculations were carried out using a $5 \times 5 \times 5$ supercell of the rock-salt crystal structure at 0~K. The same system size was used to determine defect formation energies and to compute the activation energies of point defect migration by using the nudged elastic band (NEB) method implemented in LAMMPS.
Phonon dispersions were computed using the finite-displacement method in PHONOPY with a $3\times3\times3$ supercell and a displacement amplitude of 0.01 \AA. Ensemble-averaged phonons were obtained by averaging the independently computed phonon frequencies across models evaluated on the same Brillouin-zone sampling grid.
Temperature-dependent lattice parameters were obtained from a series of isobaric-isothermal (NPT) simulations performed on the MLIP ensemble. We used a $15 \times 15 \times 15$ supercell and covered a temperature range from 300~K up to 2500~K. The Nosé–Hoover thermostat and barostat were utilized to maintain the target temperature and zero pressure, respectively. Thermostat and barostat damping parameters were set to 0.1~ps and 0.5~ps, with simulations running for a total duration of 30~ps using a timestep of 1~fs. Equilibrium lattice parameters were calculated by averaging simulation box dimensions over the final 10~ps of each simulation to ensure convergence and accurate thermodynamic averaging.

Self-diffusion coefficients for carbon and uranium atoms under the presence of defects were calculated by extracting the mean square displacements from NVE simulations held for 1~ns at a specific temperature in the range of 1800~K to 2300~K. These simulations were performed on $15 \times 15 \times 15$ supercell systems with varying stoichoimetric conditions.

\section{Results}

\subsection*{MLIP Performance}

\begin{figure}[ht!]
  \centering
  \begin{subfigure}{0.48\textwidth}
    \includegraphics[width=\textwidth]{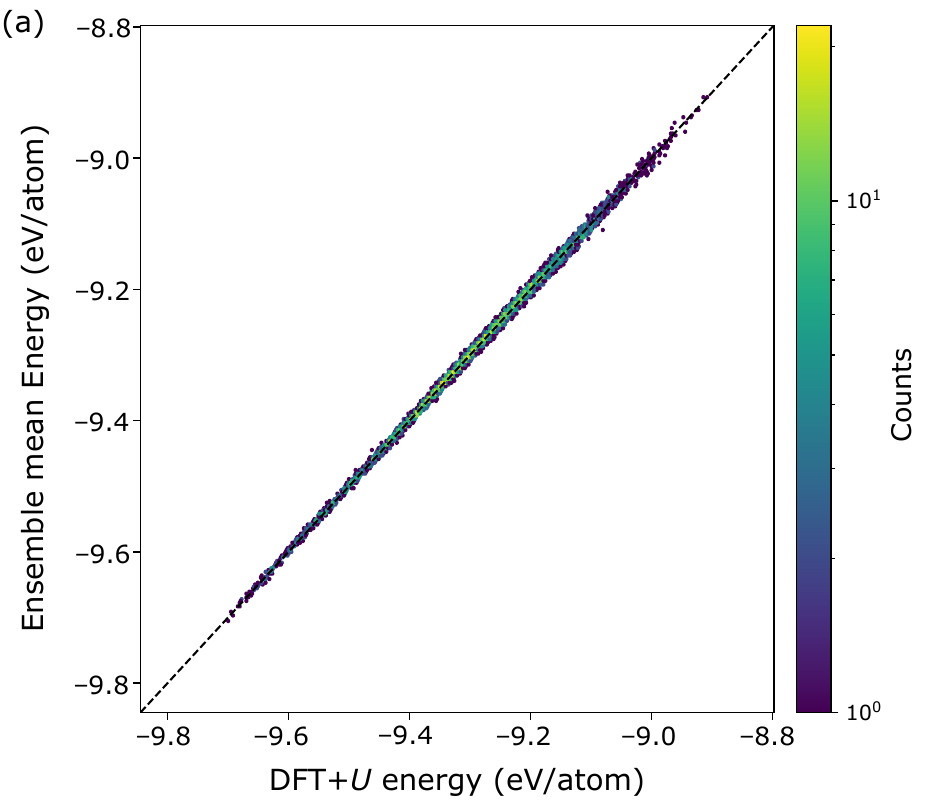}
        \phantomcaption
    \label{fig:energy}
  \end{subfigure}
  \hfill
  \begin{subfigure}{0.48\textwidth}
    \includegraphics[width=\textwidth]{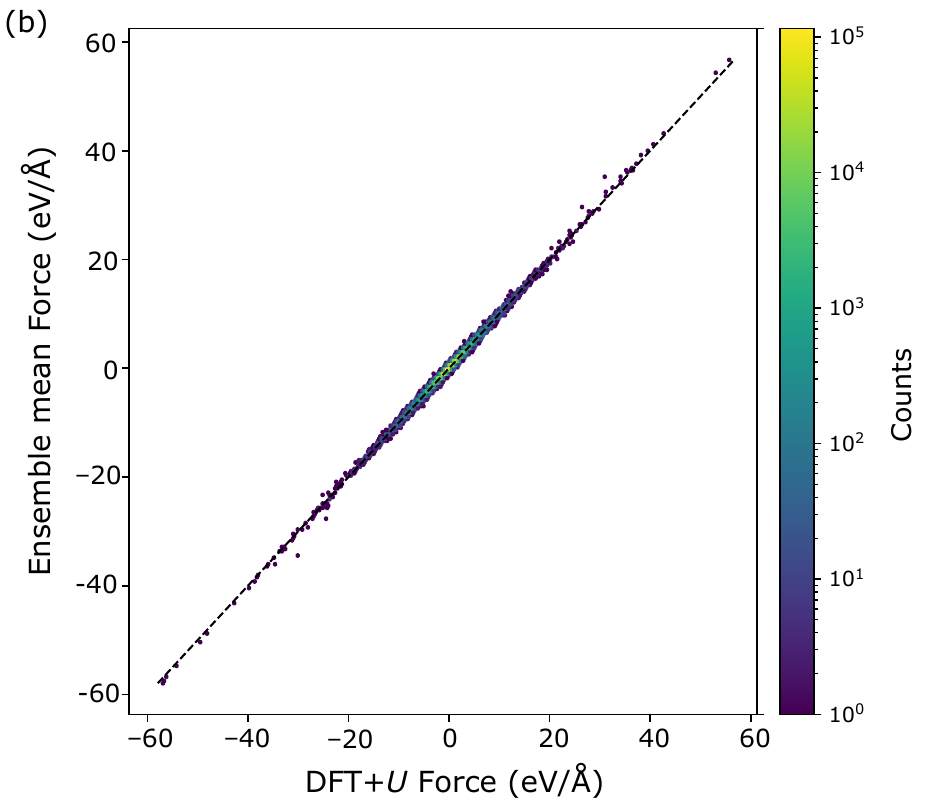}
        \phantomcaption
    \label{fig:force}
  \end{subfigure}
  \vspace{-20pt}
  \caption{Parity plots comparing ensemble HIP-NN predictions with DFT+\textit{U} reference data for (a) Energy per atom correlation plot showing ensemble-mean predictions versus DFT+\textit{U}, and (b) Force parity plot comparing ensemble-mean component forces to DFT+\textit{U} values. Colors indicate the logarithm of the point density.}
  \label{fig:parity}
\end{figure}

On the held-out test subset, the MLIP achieves a total energy root mean square error (RMSE) of 5.34 $\pm$ 0.49~meV/atom and a force RMSE of 0.203 $\pm$ 0.007~eV/\AA, where the uncertainties reflect variation across the model ensemble. As shown in Fig.~\ref{fig:parity}, the parity plots demonstrate a good agreement between the HIP-NN potential and the DFT+\textit{U} reference calculations. Overall, these values are similar to--though they are approximately a factor of two higher than--those reported for a previously developed HIP-NN MLIP for UN trained on DFT data~\cite{alzatevargas2024}, which achieved energy and force RMSEs of 2.48~meV/atom and 0.11~eV/\AA, respectively. The higher RMSEs are not completely unexpected, as DFT+\textit{U} calculations are inherently more challenging. Previous work also reported errors approximately two to four times higher when fitting an MLIP to DFT+\textit{U} data compared to DFT data for UO$_2$~\cite{Stippell2024}. By comparison, our RMSEs are significantly lower than those reported for a HIP-NN MLIP for UO$_2$ trained on DFT+\textit{U} data, which achieved energy and force RMSEs of 6.50~meV/atom and 0.44~eV/\AA, respectively~\cite{gumber_going_2026}.

Table~\ref{tab:cluster_errors} summarizes the energy and force prediction errors of the MLIP ensemble across structurally distinct clusters of uranium carbide configurations. The energy and force RMSE values across all clusters indicate a robust performance over a wide range of structural environments. Lowest energy errors are found for Frenkel defect configurations, suggesting that these structures are well represented by the model. On contrary, the largest errors are associated with neutral vacancy configurations, reflecting increased structural complexity. Other clusters shown in the table demonstrate good generalization across different subsets. See Fig.~\ref{fig:tsne} in supplementary information for a visualization of the dataset divided into the various clusters.

\begin{table}[ht]
\centering
\caption{Per-cluster energy and force errors of the HIP-NN potential relative to DFT+\textit{U} generated using the full dataset. Errors represent the ensemble average from eight independent MLIPs.}
\label{tab:cluster_errors}
\begin{ruledtabular}
\begin{tabular}{lccccc}
Cluster & $N_{\mathrm{configs}}$ & 
$E_{\mathrm{RMSE}}$ & $E_{\mathrm{MAE}}$ & 
$F_{\mathrm{RMSE}}$ & $F_{\mathrm{MAE}}$ \\
& & (meV/atom) & (meV/atom) & (eV/\AA) & (eV/\AA) \\
\hline
Distorted UC (expanded)      & 1405 & 4.84 & 3.77 & 0.190 & 0.147 \\
Distorted UC (mid-range)     & 1617 & 5.11 & 3.96 & 0.197 & 0.153 \\
Distorted UC (compressed)    & 1316 & 4.95 & 3.88 & 0.189 & 0.146 \\
Frenkel defect               &  442 & 3.08 & 2.40 & 0.130 & 0.099 \\
Bivacancy defect             & 1962 & 5.41 & 4.27 & 0.203 & 0.158 \\
\end{tabular}
\end{ruledtabular}
\end{table}

To demonstrate the robust predictive capability of the HIP-NN potential developed for UC, we evaluate some relevant thermophysical and transport properties and compare them with available experimental data. See Fig.~\ref{fig:extra} in supplementary information for additional thermophysical properties. For comparison purposes, some metrics include DFT+\textit{U} reference values and data extracted directly from the available publication for the classical potential developed by Chartier and Brutzel~\cite{Chartier2007}.\\

A first validation of the MLIP is given in Table~\ref{tab:elastict}, where we present the calculated values of lattice parameter ($a$) at 300~K and elastic constants at 0~K.
We compare the MLIP predictions with the classical potential values published in Ref.~\cite{Chartier2007}, the calculated DFT+\textit{U} values from this work, and reference experimental data at 300~K~\cite{graham1963,routbort1971,Butorin2023}. We find that the HIP-NN MLIP predicts a lower value of $C_{11}$ in comparison to the classical potential and the DFT+\textit{U} data, while achieving higher accuracy for the values of $C_{12}$ and $C_{44}$.

\begin{table}[ht!] 
\caption{\label{tab:elastict} Comparison of lattice parameter ($a$) at 300~K and UC elastic constants calculated at 0~K using the MLIP, a classical potential, DFT+\textit{U} calculations. Experimental data obtained at 300~K.}
\begin{ruledtabular}
\begin{tabular}{ccccc}
                                    & $a$ 300 K & $C_{11}$  & $C_{12}$ & $C_{44}$\\
                                    & (\AA) & (GPa) & (GPa) & (GPa) \\
                                    \hline 
HIP-NN                              & 4.980   & 298& 80& 63\\
Classical potential (Ref.~\cite{Chartier2007})& 4.952  & 322   & 104  & 39  \\
DFT+\textit{U} (This work)          & 4.985   & 333   & 73   & 48   \\
Exp. at 300~K (Refs.~\cite{graham1963,routbort1971,Butorin2023}) 
                                    & 4.956 &  314--320 & 77--79  & 64--66 
\end{tabular}
\end{ruledtabular}
\end{table}

\begin{figure}[ht!]
    \centering 
    \includegraphics[width=0.5\linewidth]{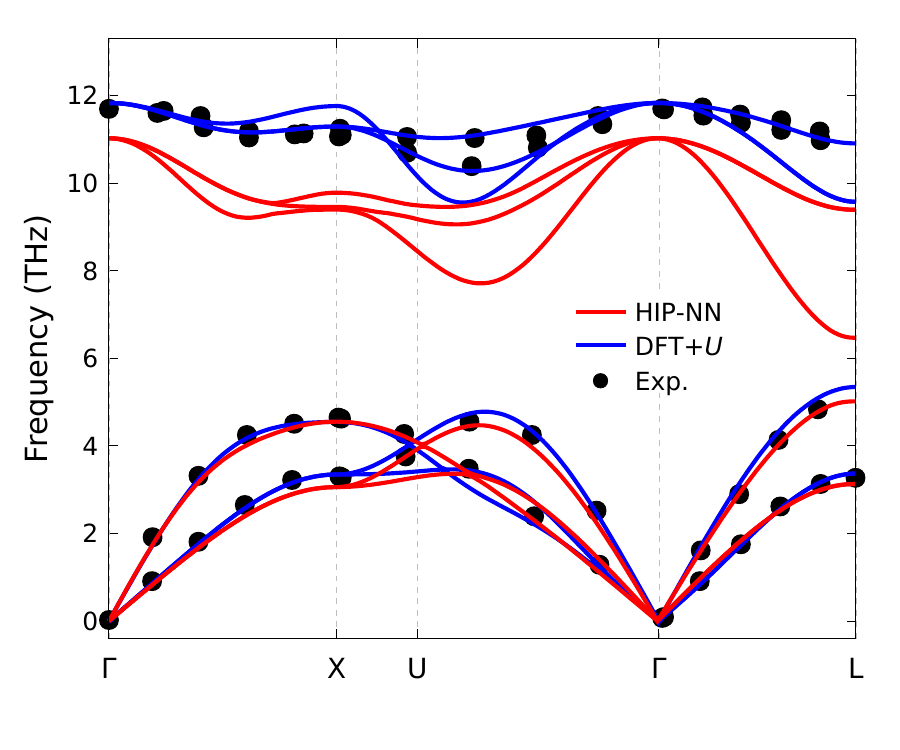} 
  \caption{\label{phononmlp}Phonon band structure as predicted by the MLIP (red). For comparison, the phonons from DFT+$U_{\mathrm{eff}}=1.25$ eV calculation on AFM structure is shown (blue). Experimental data from Ref.~\cite{jackman1986} included as black points.}
\end{figure}

Phonon band dispersions calculated at 0~K with the MLIP are shown in Fig.~\ref{phononmlp}, where DFT+$U_{\mathrm{eff}}=1.25$ eV dispersions and experimental~\cite{jackman1986} data are included for reference. The potential accurately reproduces the acoustic branches at low frequencies, which are governed by bulk elastic properties. However, the optical modes are underestimated by approximately 1 THz, indicating a softer description of the short-range U-C interactions. These interactions, and thus the optical modes, are highly sensitive to the HIP-NN tensor. As shown in the supplementary information, a slight improvement can be achieved by increasing the order of the HIP-NN tensor sensitivity (see Fig.~\ref{comp}).

\begin{figure}[ht!]
    \centering 
    \includegraphics[width=0.6\linewidth]{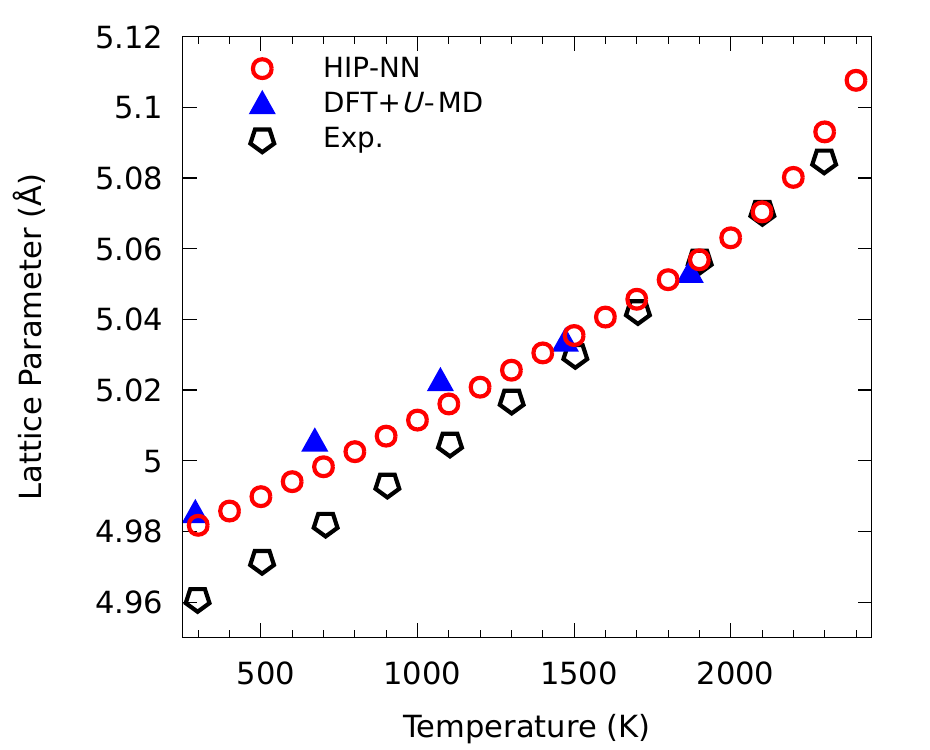} 
  \caption{\label{fig:lattice} Comparison of lattice parameter as a function of temperature obtained from HIP-NN, DFT+\textit{U}-MD, and experimental data from Ref.~\cite{meddez1964}.}
\end{figure}

To confirm the MLIP ability to simulate stoichoimetric UC in the FCC NaCl-type structure; we calculate the lattice parameter at zero pressure as a function of temperature as shown in Fig.~\ref{fig:lattice}. The MLIP is capable of performing stable MD simulations up to 2500~K, below melting temperature of 2780~K for UC~\cite{utton2009}. It should be noted that the melting point of UC is highly dependent on the carbon content and many of the uranium monocarbide high-temperature properties, including the lattice parameter, homogeneously extend up to the $\beta$-UC$_2$ composition. For comparison, Fig.~\ref{fig:lattice} includes the lattice parameter obtained from DFT+\textit{U}-MD simulations performed with the reference DFT+\textit{U} used in the training, where $U_{\mathrm{eff}}$=1.25 eV.
Our MLIP lattice parameter is in excellent agreement with the predicted DFT+\textit{U}-MD lattice parameter over the temperature range studied. However, it is known that DFT+\textit{U} often overestimates the lattice parameter for strongly correlated systems. Consequently, we find that at temperatures below 1500~K, the MLIP also overestimates the lattice parameter compared to the experimental data. This appears to be minimized as the temperature increases, where the MLIP matches the experimental data.

\subsection*{Point Defects and Diffusion}

Radiation damage in reactor conditions produces point defects and fission products, both of which significantly impact the service performance of nuclear fuels. Point defects such as vacancies serve as trapping sites for impurities and are critical to diffusion-driven processes. To evaluate the performance of our MLIP and understand the role of defects in uranium carbide, we report the formation energies of neutral point defects—including uranium and carbon Frenkel pairs and Schottky defects—in Table~\ref{tab:defect}. While the MLIP tends to overestimate formation energies relative to DFT and experiment, the overall trends are consistent. For instance, the MLIP predicts a formation energy of 4.4~eV for the carbon Frenkel pair, overestimating the experimental value of 2.2~eV~\cite{matzke1984}. It is important to interpret these comparisons with caution, as experimental measurements often involve samples with oxygen and nitrogen impurities and may not reflect perfectly. While the carbon FP value predicted by the classical potential tailored to model diffusion in UC is in closer agreement with the experimental value, the MLIP value is in close agreement with the DFT+\textit{U} value reported in this work.

\begin{table} 
\caption{\label{tab:defect} Comparison of neutral point defect formation energies for uranium Frenkel pair (FP$_\text{U}$), carbon Frenkel pair (FP$_\text{C}$), and the Schottky defect (SD) obtained from MLIP, a classical potential, DFT/DFT+\textit{U} calculations, and experiment. No value for SD was provided by Ref.~\cite{Chartier2007} for the classical potential.}
\begin{ruledtabular}
\begin{tabular}{cccc}
                    & FP$_\text{U}$ &  FP$_\text{C}$ & SD  \\ 
                    & (eV) & (eV) & (eV) \\ 
   \hline
HIP-NN                                           & 9.5& 4.4& 5.5\\
Classical potential (Ref.~\cite{Chartier2007})   & 6.8   & 1.5  & -    \\
DFT (Ref.~\cite{Ducher2011})                     & 7.6   & 3.4  & 4.3  \\
DFT+\textit{U} (This work)                       & 8.1   & 3.4  & 3.8  \\
Experiment (Ref.~\cite{matzke1984})              & - & 2.2 & - \\

\end{tabular}
\end{ruledtabular}
\end{table}

The migration barriers of uranium and carbon vacancies (vac) and interstitials (int) are presented in Table \ref{tab:neb}. We report values predicted by our MLIP, alongside data from a classical potential, DFT, DFT+\textit{U} calculations available in the literature, and experimental measurements. The experimental values were extracted from electrical resistivity measurements of quenched UC samples, as reported by Sch{\"u}le and Spindler~\cite{schule1969} and later by Matsui and Matzke~\cite{Matsui1980}. These measurements used samples intended to be stoichiometric UC, although all specimens contained measurable oxygen and nitrogen impurities.

Our MLIP predicts a migration energy of 2.41~eV for carbon vacancies and and 2.86~eV for uranium vacancies. While DFT(+\textit{U}) calculations and the classical potential report a higher barrier for carbon vacancies with respect to uranium vacancies, the experimental data follow an opposite trend. Sch{\"u}le nd Spindler reported experimental activation energies of 1.45 eV for carbon vacancy migration and 2.2 eV for uranium. Matsui and Matzke found even lower values--0.9 eV for carbon and 2.4 eV for uranium--emphasizing the variability introduced by the microstructure and chemical impurities. 
This trend in the experimental data is in alignment with the HIP-NN calculations reported in this work.

The MLIP correctly captures relative trends found in DFT and DFT+\textit{U} calculations --such as the lower barrier for interstitial diffusion versus vacancy diffusion. However, the migration energy for an uranium interstitial predicted by the MLIP is 0.66~eV, under predicting DFT+\textit{U} calculations by approximately 0.4~eV. Carbon interstitial migration energy calculated using the MLIP was found to be near zero, an error of nearly 1~eV. These low interstitial migration barriers, however, can be explained by the lack of transition-state-like configurations in the training dataset. Future refinements could improve the quantitative accuracy of interstitial activation energies by performing direct NEB sampling to include migration pathways in the training dataset. 

\begin{table}
\caption{\label{tab:neb} Activation energy for migration of uranium and carbon vacancies (vac) and interstitials (int) extracted from NEB calculations using the MLIP and compared to a classical potential, DFT and DFT+\textit{U} NEB calculations, and experimental data.}
\begin{ruledtabular}
\begin{tabular}{ccccc}
    &  C$_\text{vac}$  & U$_\text{vac}$ &  C$_\text{int}$ & U$_\text{int}$  \\
    & (eV) & (eV) & (eV) & (eV) \\
   \hline       
HIP-NN                                           & 2.41& 2.86& 0.04& 0.66\\
Classical potential (Ref.~\cite{Chartier2007})   & 4.8   & 2.2  & 1.8  & 0.9  \\
DFT (Ref.~\cite{Ducher2011})                     & 2.0   & 1.8  & 1.6  & 3.0  \\
DFT+\textit{U} (Ref.~\cite{Huang2020})           & 2.5   & 1.9  & 0.98 & 1.09 \\
Exp. (Sch{\"u}le and Spindler~\cite{schule1969}) & 1.45  & 2.2  &  -   &  -   \\
Exp. (Matsui and Matzke~\cite{Matsui1980})       & 0.9   & 2.4  &  -   &  -   \\
\end{tabular}
\end{ruledtabular}
\end{table}

We further extend the prediction of diffusion mechanisms by computing the activation energies from Arrhenius fits to the MD diffusion data and explore relevant experimental data. One should note that direct comparison between MLIP predictions and experiments remains challenging. Predicting MLIP diffusion constants requires running MD simulations at relatively high temperatures to reach the diffusive regime. Comparison with experiment at high temperatures is especially problematic, since UC could exist as a two-phase material (e.g., U + UC or UC + UC$_2$), affecting the measured transport properties. 
Consequently, experimental activation energies from self-diffusion vary widely --from 1.9 to 4.1 eV for carbon~\cite{chubb1964, lee1968, krakowski1969, makino1972}, and from 3.6 to 5.4 eV for uranium~\cite{schule1969,bentle1968}, strongly depending on composition (carbon content) and microstructure. 
Our MLIP predicts that uranium diffuses significantly more slowly than carbon. This trend is consistent with experiments, which report that carbon diffusion in UC is approximately $10^2$–$10^3$ times faster than uranium diffusion~\cite{schule1969,bentle1968}. As a result, uranium diffusion in pure UC is not accessible within the time scales of our MD simulations. Instead, we report uranium diffusion for systems containing uranium defects, where the mobility is sufficiently enhanced.

\begin{figure}[ht!]
  \centering
  \begin{subfigure}{0.55\textwidth}
    \includegraphics[width=\textwidth]{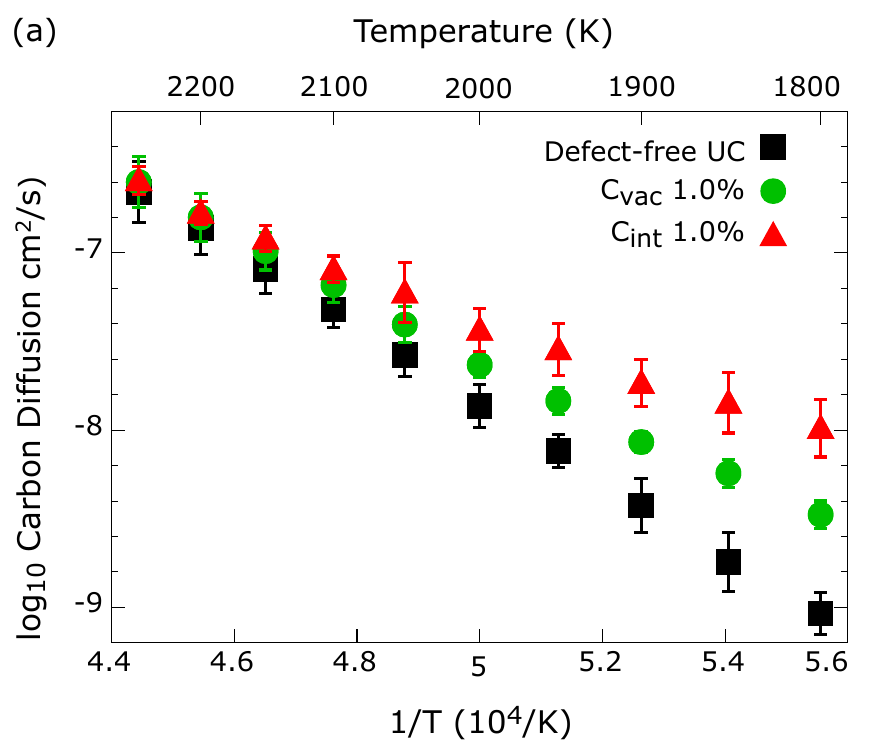}
        \phantomcaption
    \label{fig:cdiff}
  \end{subfigure}
  \hfill
  \begin{subfigure}{0.55\textwidth}
    \includegraphics[width=\textwidth]{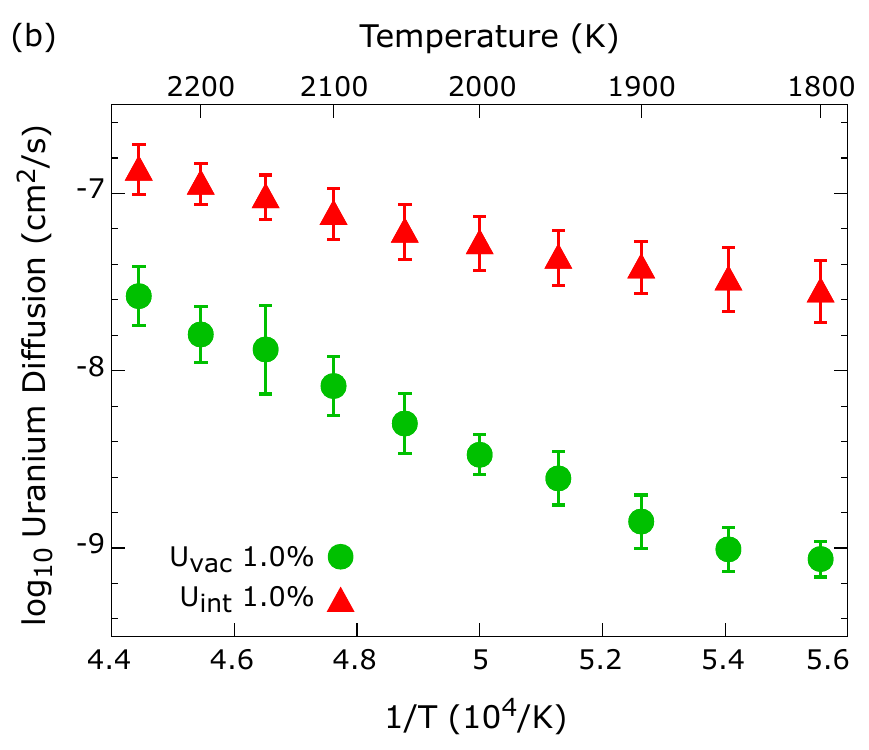}
        \phantomcaption
    \label{fig:udiff}
  \end{subfigure}
  \vspace{-20pt}
  \caption{Self-diffusion coefficients as a function of temperature for (a) Carbon in defect-free UC, carbon-deficient sample (C$_\text{vac}$), and a carbon-rich sample (C$_\text{int}$), and (b) Uranium in an uranium-deficient system (U$_\text{vac}$) and an uranium-rich (U$_\text{int}$).}
  \label{fig:diff}
\end{figure}

Figure~\ref{fig:cdiff} presents the self-diffusion coefficients of carbon in UC as a function of inverse temperature for three systems: a defect-free (stoichiometric) lattice, a carbon-deficient system with 1.0\% vacancies, and a carbon-rich system with 1.0\% interstitials. 

In the stoichiometric case, the MLIP predicts an activation energy of 4.3~eV, almost 2~eV higher than experimental values. For instance, Lee and Barrett~\cite{lee1968} reported an activation energy of 2.7~eV between 1539--1957~K, while Makino et al.~\cite{makino1972} found a value of 2.3~eV between 1600--2200~K. These discrepancies likely stem from the idealized nature of the simulations: the absence of native or thermally activated defects in the pristine simulation cell inhibits diffusion, leading to artificially high barriers.
The presence of carbon vacancies lowers the activation energy to 3.2~eV, while interstitials reduce it further to 2.7~eV. This trend aligns with experimental observations that carbon diffusivity in UC is highly composition-dependent. The simulation results suggest that interstitial-mediated diffusion is the dominant path for carbon mobility under carbon-rich conditions in correlation with experimental findings. Several experiments~\cite{chubb1964, lee1968} report activation energies ranging from 2.7 to 3.9~eV on carbon-deficient samples, while carbon-rich structures have energies from 1.8--2.3 eV. While any conclusions regarding these values should be taken with care when comparing to experiments, the MLIP captures values within these ranges or slightly higher.

The summary of carbon activation energies shown in Table~\ref{tab:energies_d}, demonstrates that the MLIP accurately captures the relative trends in activation energy and the effect of stoichiometry, even if the absolute barriers for vacancy diffusion remain slightly overestimated.

\begin{table}[ht]
\centering
\caption{Comparison of MLIP-predicted activation energies (\textit{E$_a$}) with experimental data for carbon diffusion in UC across different compositions.}
\begin{ruledtabular}
\begin{tabular}{lcccccc}
Defect & MLIP \textit{E$_a$}  &  Exp. Composition & Exp. \textit{E$_a$} & Exp. T & Refs. \\
& (eV) &  & (eV) & (K) &  \\
\hline
Pure UC (defect-free)    & 4.3 & Stoichiometric      & 2.3--2.7 & 1500--2300 & \cite{lee1968, makino1972} \\
C vacancy                        & 3.3 & Carbon-deficient  & 2.7--3.9 & 1300--2100 & \cite{lee1968, chubb1964, krakowski1969} \\
C interstitial                    & 2.5 & Carbon-rich         & 1.8--2.3  & 1470--2300 & \cite{lee1968, makino1972} \\
\end{tabular}
\label{tab:energies_d}
\end{ruledtabular}
\end{table}

Figure~\ref{fig:udiff} shows the self-diffusion coefficients of uranium in UC for systems containing either uranium vacancies or interstitials. Under stoichiometric or uranium-deficient conditions, diffusion is extremely slow. When diffusion is facilitated by vacancies, the MLIP predicts an activation energy of 2.7~eV for uranium diffusion, matching the calculated NEB migration barriers for vacancy defects.
When uranium interstitials are introduced --uranium-rich (C-deficient) compositions--the predicted activation energy drops significantly to 1.2~eV,  validating the ability of the MLIP to capture defect-driven transport trends. 

\section{Conclusions}
We present the first MLIP specifically developed for uranium monocarbide, trained on high-quality  DFT+\textit{U} data and validated across critical material properties. The MLIP demonstrates robust predictive capabilities, accurately reproducing lattice parameters, elastic constants, defect energetics, and diffusion mechanisms across a range of thermodynamic conditions. While some activation energies for defect diffusion are moderately overestimated, relative trends remain consistent with experimental observations, particularly emphasizing the importance of defect-driven transport. This MLIP represents a significant step toward performing realistic atomistic simulations, as it achieves greater accuracy than classical potentials with lower costs than DFT calculations. This MLIP can lead to a better understanding of UC in reactor conditions and help with qualification of UC as an alternative nuclear fuel candidate.

\begin{acknowledgments}
The authors are grateful for insightful discussions with Benjamin Nebgen and Nicholas Lubbers regarding active learning and machine learning potentials. Los Alamos National Laboratory (LANL), an affirmative action/equal opportunity employer, is operated by Triad National Security LLC, for the National Nuclear Security Administration of the U.S. Department of Energy under Contract No. 89233218CNA000001. The research presented in this work was supported by the Laboratory Directed Research and Development (LDRD) program via project 20220053DR at LANL. We also appreciate the resources provided by the LANL Institutional Computing (IC) program. R.A.M. acknowledges the Oak Ridge Leadership Computing Facility at the Oak Ridge National Laboratory, which is supported by the Office of Science of the U.S. Department of Energy under Contract No. DE-AC05-00OR22725.
\end{acknowledgments}

\newpage
\begin{center}
    \Large{Supplementary Information for:\\}
\large{Atomistic modeling of uranium monocarbide with a machine learning interatomic potential.\\}
\end{center}

\subsection*{t-sne visualization of clusters}

\renewcommand{\thefigure}{S\arabic{figure}}
\setcounter{figure}{0}

\renewcommand{\thetable}{S\arabic{table}}
\setcounter{table}{0}

\begin{figure}[h!]
    \centering 
    \includegraphics[width=0.7\linewidth]{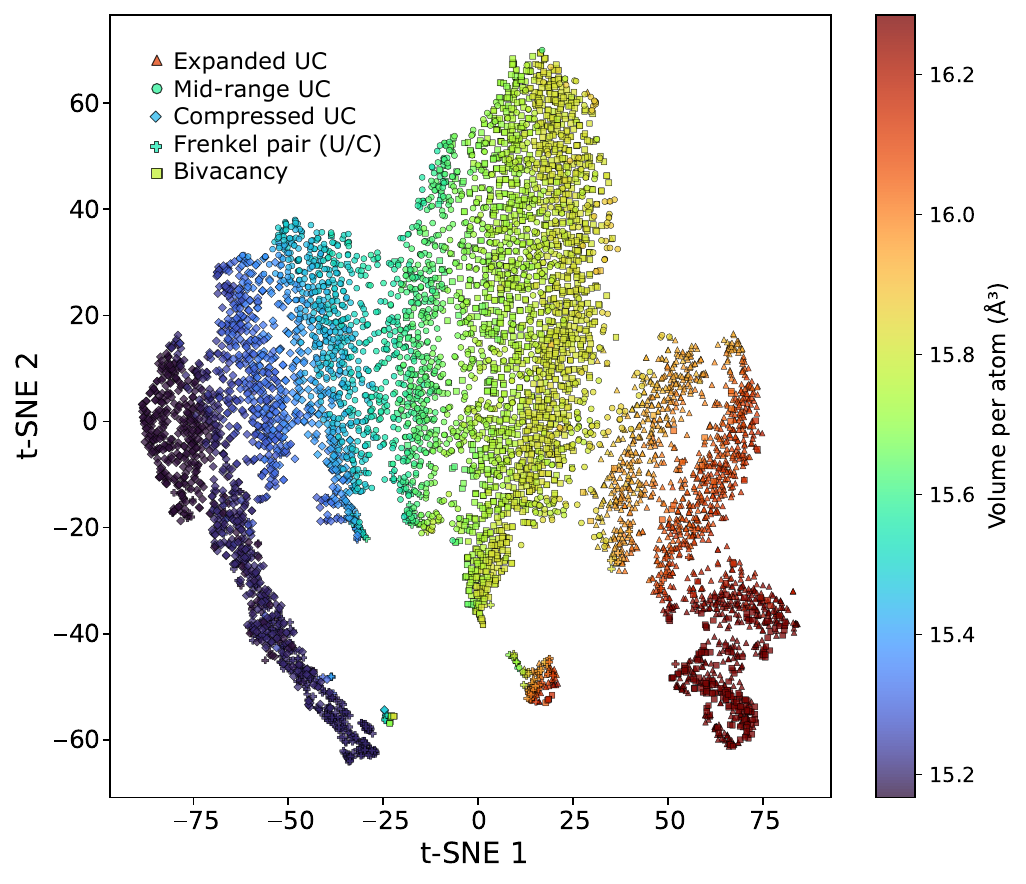} 
  \caption{\label{fig:tsne} Two-dimensional t-SNE visualization of the UC dataset after defect-based classification identified using a lattice-based analysis: stoichiometric distorted (expanded, mid-range, and compressed), Frenkel pairs, and neutral vacancy configurations. Structures are color coded based on volume per atom.}
  %The t-SNE embedding is used solely for visualization of structural relationships; defect classification and clustering were performed using explicit physical criteria.}
\end{figure}

Figure~\ref{fig:tsne} presents a t-distributed stochastic neighbor embedding (t-SNE) visualization of the dataset, where points are colored based on the volume per atom. The data display a clear clustering between structure and volumetric strain. Atomic configurations are represented using a global structural descriptor based on interatomic distance distributions. Defect-containing configurations, including Frenkel defects and bivacancies, span a broader range of volumes, reflecting the local strain and lattice distortions introduced by point defects.

\clearpage
\newpage

\subsection*{Temperature-dependent heat capacity and bulk modulus}

\begin{figure}[h!]
  \centering
  \begin{subfigure}{0.5\textwidth}
    \includegraphics[width=\textwidth]{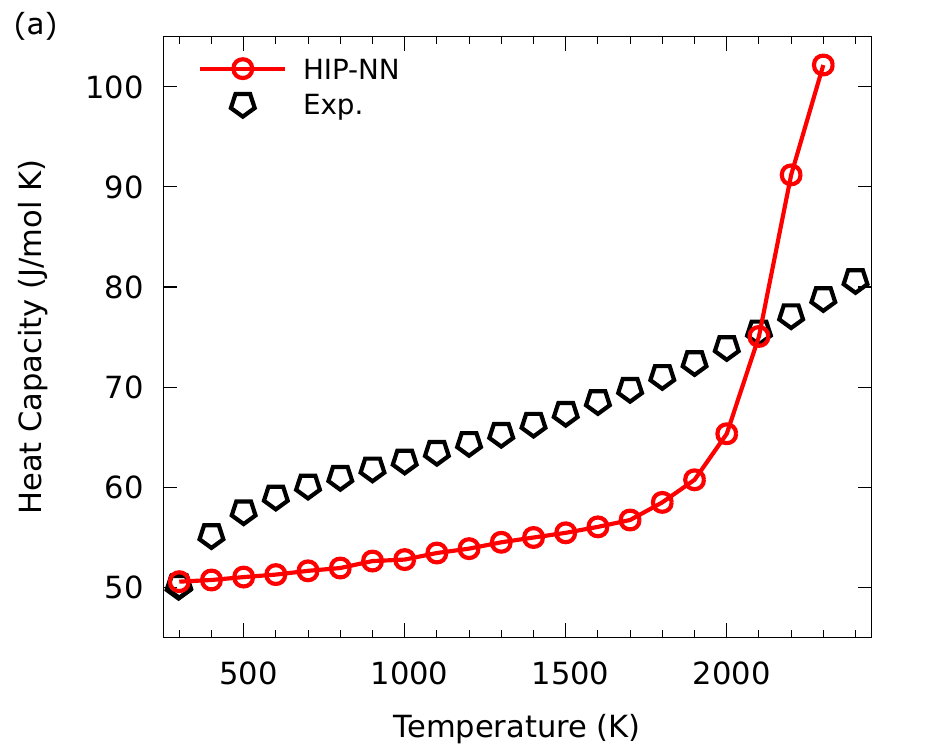}
        \phantomcaption
    \label{fig:heat}
  \end{subfigure}
  \hspace{-20pt}
  \begin{subfigure}{0.5\textwidth}
    \includegraphics[width=\textwidth]{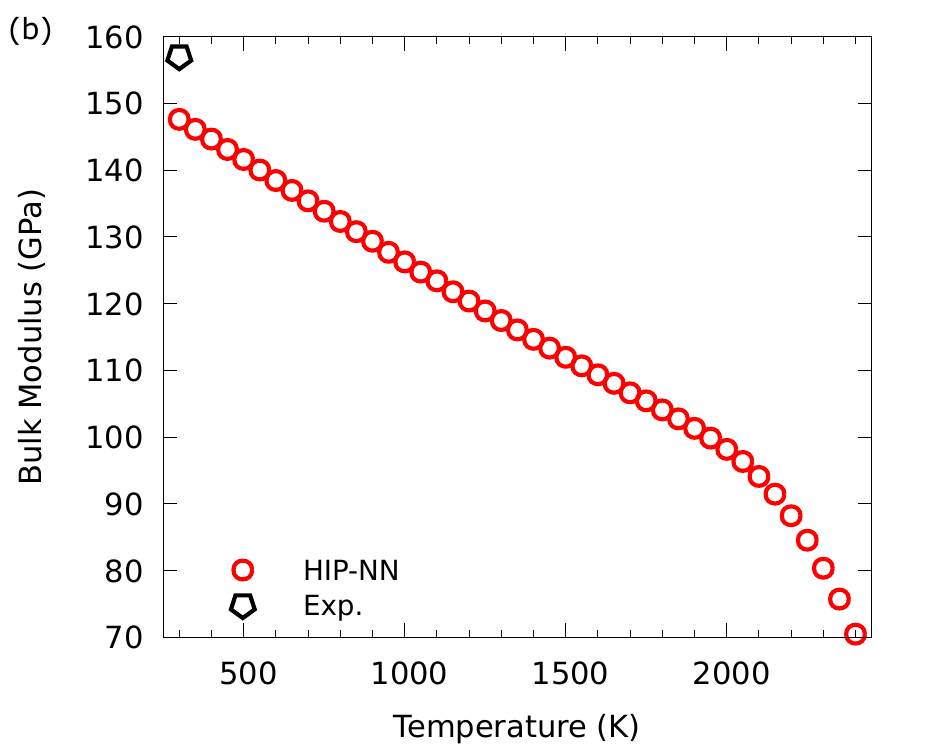}
        \phantomcaption
    \label{fig:bulk}
  \end{subfigure}
  \vspace{-25pt}
  \caption{MLIP predicted thermophysical properties as function of temperature. (a) Specific heat capacity with experimental data from Ref.~\textcolor{blue}{[1]}, and (b) Bulk moduli with experimental data from Ref.~\textcolor{blue}{[2]}.}
  \label{fig:extra}
\end{figure}

Fig.~\ref{fig:heat} presents the specific heat capacity at constant pressure. Fig.~\ref{fig:bulk} presents the bulk moduli as a function of temperature. Both figures compare the HIP-NN MLIP predictions with available experimental data.

Since MD calculations only capture the vibrational component of the heat capacity--excluding electronic (quantum) and magnetic contributions-- there is a large discrepancy with the experimental-inferred heat capacity~\textcolor{blue}{[1]}. As depicted in Fig.~\ref{fig:heat}, the MLIP shows a steady increase of the heat capacity with a sudden increase after 2000~K, that is not reflected in the experimental data.
The bulk moduli calculated at zero-pressure, correctly predicts the expected inverse relation between bulk modulus and lattice expansion (see Fig.~\ref{fig:bulk}). The MLIP bulk modulus at 300~K is lower than the experimental data, which is around 156--158~GPa for stoichoimetric UC~\textcolor{blue}{[2]}.

\clearpage
\newpage

\subsection*{Potential performance with increased tensor sensitivity}

Constructing an MLIP often requires tedious hyperparameter tuning. One key hyperparameter for HIP-NN is the order of tensor sensitivity $({\ell_\text{max}})$. Although higher tensor sensitivity often results in lower training errors, this comes at a greater computational cost for running MD and potentially a higher risk of overfitting. Thus, 
in this section, we investigate the impact of increasing the HIP-NN tensor sensitivity on the model performance, in particular, on the phonon dispersion, elastic constants, and defect energetics.
For this purpose, we train a model with quadrupole tensor sensitivity (${\ell_\text{max}}=2$, HipnnQuad) and compare with the model presented in the main text that utilizes vector tensor sensitivity (${\ell_\text{max}}=1$, HipnnVec). For both models, we set the cusp regularization to 0.01 as this value is essential to ensure mechanical stability. Specifically, a HIP-NN ${\ell_\text{max}}=2$ model with a lower cusp regularization ($1e^{-6}$) exhibited unphysical elastic behavior.

\begin{table}[h]
\centering
\caption{Elastic constants (GPa) obtained from MLIP models compared to DFT+$U$ and a classical potential.}
\label{tab:elastic_constants}
\begin{ruledtabular}
\begin{tabular}{lccc}
Model & $C_{11}$ & $C_{12}$ & $C_{44}$ \\
\hline
HipnnVec  & 296.6 & 84.2 & 63.5 \\
HipnnQuad  & 297.8 & 91.4 & 28.4 \\
DFT+$U$ & 333 & 73 & 48 \\
\end{tabular}
\end{ruledtabular}
\end{table}

By comparing the elastic constants (Table~\ref{tab:elastic_constants}) for both MLIPs, we find that the $C_{11}$ and $C_{12}$ values are quite similar for HipnnVec and HipnnQuad. However, the HipnnQuad model substantially underestimates the shear modulus $C_{44}$, indicating an overly soft response to shear deformations. 

\begin{figure}[ht!]
    \centering 
    \includegraphics[width=0.5\linewidth]{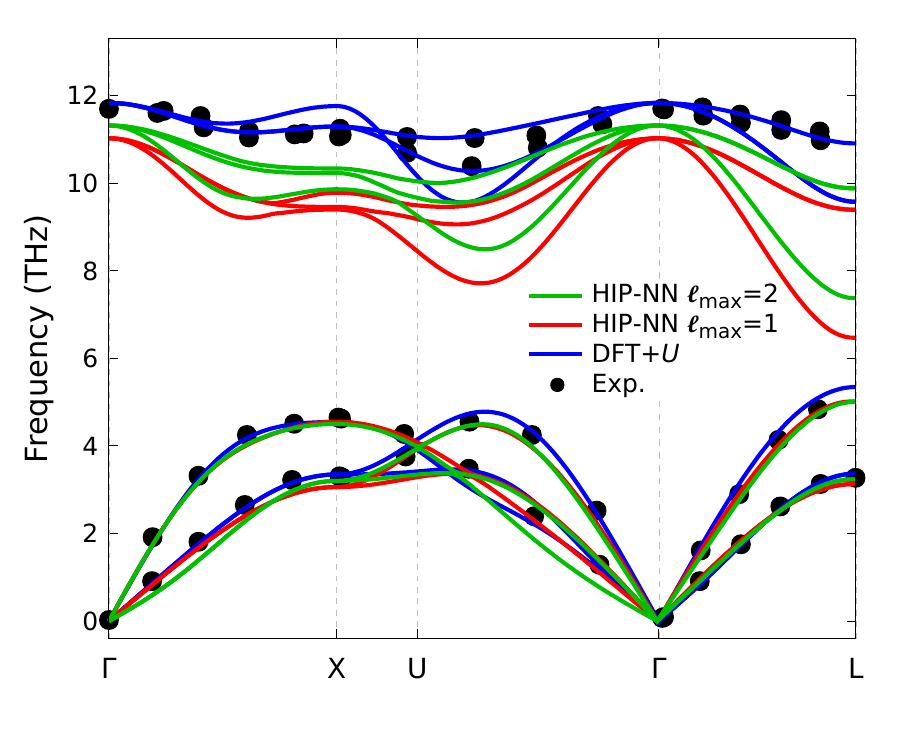} 
  \caption{\label{comp}Phonon band structures predicted by MLIPs using two different tensor sensitivities of ${\ell_\text{max}}=1$ (HipnnVec)--as shown in the main text-- and ${\ell_\text{max}}=2$ (HipnnQuad). DFT+$U_{\mathrm{eff}}=1.25$ eV calculation on AFM structure and experimental data from Ref.~\cite{jackman1986} included for reference.}
\end{figure}

The phonon dispersions, shown in Fig.~\ref{comp}, reveal that the quadrupole model (HipnnQuad) reproduces optical phonon branches more closely than the vector model (HipnnVec) along high-symmetry paths. The observed trade-off between optical phonon accuracy and elastic stability underscores that phonon agreement away from $\Gamma$ is not sufficient to guarantee physically correct elastic behavior.

Table~\ref{tab:defect_energies} summarizes Schottky defect and Frenkel energies for carbon and uranium predicted by the two MLIPs, compared against DFT+$U$. The HipnnQuad model reproduces defect formation energies more closely than the HipnnVec model, by approximately 0.3 to 0.7~eV. Defect formation energies are dominated by highly localized distortions, which are well captured by higher-order angular sensitivity channels available in the quadrupole model.

\begin{table}[h]
\centering
\caption{Point-defect formation energies (eV) predicted by MLIP models compared to reference data.}
\label{tab:defect_energies}
\begin{ruledtabular}
\begin{tabular}{lccc}
Model &  SD & FP$_\text{C}$ & FP$_\text{U}$ \\
\hline
HipnnVec  & 5.58 & 4.41 & 9.60 \\
HipnnQuad & 5.22 & 4.15 & 8.84 \\
DFT+$U$ & 4.3 & 3.4 & 8.1 \\
\end{tabular}
\end{ruledtabular}
\end{table}

Based on the combined assessment of elastic constants, phonon stability and defect energetics the HipnnVec model provides a more balanced consistent description of uranium carbide. While HipnnQuad yields improved optical phonon agreement and defect formation energies, HipnnQuad is extremely deficient in predicting the shear response. Our decision to utilize HipnnVec as a general-purpose production potential is motivated by the similar accuracy to HipnnQuad and the lower computational cost for running MD.

\subsection*{References}
\footnotesize{
\noindent [1] G. Vasudevamurthy and A. T. Nelson, Uranium carbide properties for advanced fuel modeling--A review, Journal of Nuclear Materials \textbf{558}, 153145 (2022).\\}
\footnotesize{[2] J. Routbort, Adiabatic elastic constants of uranium monocarbide, Journal of Nuclear Materials \textbf{40}, 17 (1971)}

\newpage
\bibliography{export}

\end{document}